\documentclass[preprint]{aastex632}

\newcommand{\foraas}[1]{}

\begin{document}

\newcommand{\TotalObjects}{2,299,726}
\newcommand{\TotalDiaObjects}{1,089,818}
\newcommand{\fields}{7}
\newcommand{\TotalObjectForcedSources}{268,796,943}
\newcommand{\TotalDiaObjectForcedSources}{196,911,566}
\newcommand{\TotalDiaSources}{3,086,404}

\title{Variability-finding in Rubin Data Preview 1 with LSDB}

\author[0000-0001-7179-7406]{Konstantin Malanchev}
\affil{LSST Interdisciplinary Network for Collaboration and Computing Frameworks, 933 N. Cherry Avenue, Tucson AZ 85721, USA}
\foraas{\email[show]{malanchev@cmu.edu}}
\email{malanchev@cmu.edu}
\affil{The McWilliams Center for Cosmology \& Astrophysics, Department of Physics, Carnegie Mellon University, Pittsburgh, PA 15213, USA}

\author[0000-0002-1074-2900]{Melissa DeLucchi}
\foraas{\email{delucchi@andrew.cmu.edu}}
\affil{LSST Interdisciplinary Network for Collaboration and Computing Frameworks, 933 N. Cherry Avenue, Tucson AZ 85721, USA}
\affil{The McWilliams Center for Cosmology \& Astrophysics, Department of Physics, Carnegie Mellon University, Pittsburgh, PA 15213, USA}

\author[0000-0003-3287-5250]{Neven Caplar}
\affil{LSST Interdisciplinary Network for Collaboration and Computing Frameworks, 933 N. Cherry Avenue, Tucson AZ 85721, USA}
\affil{DiRAC Institute and the Department of Astronomy, University of Washington, 3910 15th Ave NE, Seattle, WA 98195, USA}
\foraas{\email{ncaplar@uw.edu}}

\author[0000-0002-8676-1622]{Alex I. Malz}
\affil{LSST Interdisciplinary Network for Collaboration and Computing Frameworks, 933 N. Cherry Avenue, Tucson AZ 85721, USA}
\affil{The McWilliams Center for Cosmology \& Astrophysics, Department of Physics, Carnegie Mellon University, Pittsburgh, PA 15213, USA}
\foraas{\email{aimalz@nyu.edu}}

\author{Anastasia	Alexov}
\affil{Association of Universities for Research in Astronomy, 1331 Pennsylvania Ave. NW, Washington, DC 20004, USA}
\affiliation{NSF NOIRLab/NSF–DOE Vera C. Rubin Observatory HQ, 950 N. Cherry Ave., Tucson, AZ 85719, USA}
\foraas{\email{aalexov@lsst.org}}

\author[0000-0002-5592-023X]{Eric Aubourg} 
\affiliation{Université Paris Cité, CNRS, CEA, AstroParticule et Cosmologie, F-75013, Paris, France}
\foraas{\email{aubourg@in2p3.fr}}

\author[0000-0001-9037-6981]{Amanda E Bauer}
\affil{Yerkes Observatory, 373 West Geneva Street, Williams Bay, WI 53191, USA}
\foraas{\email{abauer@yerkesobservatory.org}}

\author[0009-0003-1791-8707]{Wilson Beebe}
\affil{LSST Interdisciplinary Network for Collaboration and Computing Frameworks, 933 N. Cherry Avenue, Tucson AZ 85721, USA}
\affil{DiRAC Institute and the Department of Astronomy, University of Washington, 3910 15th Ave NE, Seattle, WA 98195, USA}
\foraas{\email{wbeebe@uw.edu}}

\author[0000-0001-8018-5348]{Eric~C. Bellm}
\affil{DiRAC Institute and the Department of Astronomy, University of Washington, 3910 15th Ave NE, Seattle, WA 98195, USA}
\foraas{\email{ecbellm@uw.edu}}

\author[0000-0002-8622-4237]{Robert David Blum}
\affiliation{NSF NOIRLab/NSF–DOE Vera C. Rubin Observatory HQ, 950 N. Cherry Ave., Tucson, AZ 85719, USA}
\foraas{\email{bob.blum@noirlab.edu}}

\author[0009-0009-7822-7110]{Doug Branton}
\affil{LSST Interdisciplinary Network for Collaboration and Computing Frameworks, 933 N. Cherry Avenue, Tucson AZ 85721, USA}
\affil{DiRAC Institute and the Department of Astronomy, University of Washington, 3910 15th Ave NE, Seattle, WA 98195, USA}
\foraas{\email{brantd@uw.edu}}

\author[0009-0007-9870-9032]{Sandro Campos}
\affil{LSST Interdisciplinary Network for Collaboration and Computing Frameworks, 933 N. Cherry Avenue, Tucson AZ 85721, USA}
\affil{The McWilliams Center for Cosmology \& Astrophysics, Department of Physics, Carnegie Mellon University, Pittsburgh, PA 15213, USA}
\foraas{\email{scampos@andrew.cmu.edu}}

\author{Daniel Calabrese}
\affil{Association of Universities for Research in Astronomy, 1331 Pennsylvania Ave. NW, Washington, DC 20004, USA}
\affiliation{NSF NOIRLab/NSF–DOE Vera C. Rubin Observatory HQ, 950 N. Cherry Ave., Tucson, AZ 85719, USA}
\foraas{\email{dcalabrese@lsst.org}}

\author[0000-0002-3936-9628]{Jeffrey~L. Carlin} 
\affiliation{NSF NOIRLab/NSF–DOE Vera C. Rubin Observatory HQ, 950 N. Cherry Ave., Tucson, AZ 85719, USA}
\foraas{\email{jcarlin@lsst.org}}

\author[0000-0003-1680-1884]{Yumi Choi}
\affil{NSF National Optical-Infrared Astronomy Research Laboratory, 950 North Cherry Avenue, Tucson, AZ 85719, USA}
\foraas{\email{yumi.choi@noirlab.edu}}

\author[0000-0001-5576-8189]{Andrew Connolly}
\affil{LSST Interdisciplinary Network for Collaboration and Computing Frameworks, 933 N. Cherry Avenue, Tucson AZ 85721, USA}
\affil{DiRAC Institute and the Department of Astronomy, University of Washington, 3910 15th Ave NE, Seattle, WA 98195, USA}
\foraas{\email{ajc@uw.edu}}

\author[0000-0002-5995-9692]{Mi~Dai}
\affil{LSST Interdisciplinary Network for Collaboration and Computing Frameworks, 933 N. Cherry Avenue, Tucson AZ 85721, USA}
\affil{
        Pittsburgh Particle Physics, Astrophysics, and Cosmology Center (PITT PACC).
        Physics and Astronomy Department, University of Pittsburgh,
        Pittsburgh, PA 15260, USA
}
\foraas{\email{mi.dai@pitt.edu}}

\author{Philip~N.	Daly}
\affil{Steward Observatory and Department of Astronomy, University of Arizona, 933 N. Cherry Avenue, Tucson, AZ 85721, USA}
\foraas{\email{pndaly@arizona.edu}}

\author{Felipe Daruich}
\affil{NSF NOIRLab/NSF–DOE Vera C. Rubin Observatory HQ, 950 N. Cherry Ave., Tucson, AZ 85719, USA}
\foraas{\email{fdaruich@lsst.org}}

\author[0009-0004-4351-5968]{Guillaume Daubard}
\affil{Sorbonne Université, CNRS/IN2P3, Laboratoire de Physique Nucléaire et de Hautes Energies (LPNHE), FR-75005 Paris, France}
\foraas{\email{guillaume.daubard@lpnhe.in2p3.fr}}

\author{Francisco Delgado}
\affil{Princeton Plasma Physics Laboratory, Princeton University, Princeton, New Jersey 08543, USA}
\foraas{\email[]{femedebe@gmail.com}}

\author[0000-0002-7790-9971]{Holger Drass}
\affil{NSF NOIRLab/NSF–DOE Vera C. Rubin Observatory HQ, 950 N. Cherry Ave., Tucson, AZ 85719, USA}
\foraas{\email{hdrass@lsst.org}}

\author[0000-0003-0042-6936]{Gloria Fonseca~Alvarez}
\affil{NSF NOIRLab/NSF–DOE Vera C. Rubin Observatory HQ, 950 N. Cherry Ave., Tucson, AZ 85719, USA}
\foraas{\email{gloria.fonseca@noirlab.edu}}

\author[0000-0001-6728-1423]{Emmanuel Gangler}
\affil{Université Clermont Auvergne, CNRS, LPCA, 63000 Clermont-Ferrand, France}
\foraas{\email{emmanuel.gangler@clermont.in2p3.fr}}

\author[0000-0003-0800-8755]{Leanne~P. Guy}
\affil{NSF NOIRLab/NSF–DOE Vera C. Rubin Observatory HQ, 950 N. Cherry Ave., Tucson, AZ 85719, USA}
\foraas{\email{leanne@lsst.org}}

\author[0000-0001-5250-2633]{\v{Z}eljko Ivezi\'{c}}
\affil{DiRAC Institute and the Department of Astronomy, University of Washington, 3910 15th Ave NE, Seattle, WA 98195, USA}
\foraas{\email{ivezic@uw.edu}}

\author{Fabrice Jammes}
\affil{Université Clermont Auvergne, CNRS, LPCA, 63000 Clermont-Ferrand, France}
\foraas{\email{fabrice.jammes@clermont.in2p3.fr}}

\author[0000-0002-1578-6582]{Buell~T. Jannuzi}
\affil{Steward Observatory and Department of Astronomy, University of Arizona, 933 N. Cherry Avenue, Tucson, AZ 85721, USA}
\foraas{\email{buelljannuzi@arizona.edu}}

\author[0000-0001-5982-167X]{Tim Jenness}
\affil{NSF NOIRLab/NSF–DOE Vera C. Rubin Observatory HQ, 950 N. Cherry Ave., Tucson, AZ 85719, USA}
\foraas{\email{tjenness@lsst.org}}

\author{David Jimenez}
\affiliation{NSF NOIRLab/NSF–DOE Vera C. Rubin Observatory HQ, 950 N. Cherry Ave., Tucson, AZ 85719, USA}
\foraas{\email{Djimenez@lsst.org}}

\author[0009-0006-2411-723X]{Derek~T. Jones}
\affil{LSST Interdisciplinary Network for Collaboration and Computing Frameworks, 933 N. Cherry Avenue, Tucson AZ 85721, USA}
\affil{DiRAC Institute and the Department of Astronomy, University of Washington, 3910 15th Ave NE, Seattle, WA 98195, USA}
\foraas{\email{dtj1s@uw.edu}}

\author[0000-0003-1996-9252]{Mario Jurić}
\affil{DiRAC Institute and the Department of Astronomy, University of Washington, 3910 15th Ave NE, Seattle, WA 98195, USA}
\foraas{\email{mjuric@uw.edu}}

\author[0000-0003-4833-9137]{Steven~M. Kahn}
\affil{Departments of Physics and Astronomy, University of California, Berkeley, CA  94720}
\foraas{\email{stevkahn@berkeley.edu}}

\author[0000-0002-5261-5803]{Yijung Kang}
\affiliation{NSF NOIRLab/NSF–DOE Vera C. Rubin Observatory HQ, 950 N. Cherry Ave., Tucson, AZ 85719, USA}
\affil{SLAC National Accelerator Laboratory, Menlo Park, CA 94025, USA}
\foraas{\email{ykang@slac.stanford.edu}}

\author[0000-0001-8783-6529]{Arun Kannawadi}
\affil{Department of Physics, Duke University, Durham, NC 27708, USA}
\foraas{\email{arun.kannawadi@duke.edu}}

\author[0009-0009-2281-7031]{Jeremy Kubica}
\affil{LSST Interdisciplinary Network for Collaboration and Computing Frameworks, 933 N. Cherry Avenue, Tucson AZ 85721, USA}
\affil{The McWilliams Center for Cosmology \& Astrophysics, Department of Physics, Carnegie Mellon University, Pittsburgh, PA 15213, USA}
\foraas{\email{jkubica@andrew.cmu.edu}}

\author[0000-0001-7178-8868]{Laurent Le~Guillou}
\affil{Sorbonne Université, CNRS/IN2P3, Laboratoire de Physique Nucléaire et de Hautes Energies (LPNHE), FR-75005 Paris, France}
\foraas{\email{llg@lpnhe.in2p3.fr}}

\author[0000-0001-5028-146X]{Olivia Lynn}
\affil{LSST Interdisciplinary Network for Collaboration and Computing Frameworks, 933 N. Cherry Avenue, Tucson AZ 85721, USA}
\affil{The McWilliams Center for Cosmology \& Astrophysics, Department of Physics, Carnegie Mellon University, Pittsburgh, PA 15213, USA}
\foraas{\email{olynn@andrew.cmu.edu}}

\author[0000-0003-2271-1527]{Rachel Mandelbaum}
\affil{LSST Interdisciplinary Network for Collaboration and Computing Frameworks, 933 N. Cherry Avenue, Tucson AZ 85721, USA}
\affil{The McWilliams Center for Cosmology \& Astrophysics, Department of Physics, Carnegie Mellon University, Pittsburgh, PA 15213, USA}
\foraas{\email{rmandelb@andrew.cmu.edu}}

\author[0000-0001-8716-6561]{Marc Moniez}
\affil{Universite Paris-Saclay, CNRS/IN2P3, IJCLab, 91405 Orsay, France}
\foraas{\email{moniez@ijclab.in2p3.fr}}

\author[0000-0001-9419-3947]{Christopher~B. Morrison}
\affil{Allen Institute for Brain Science, Seattle, WA 98103, USA}
\foraas{\email{morrison.chrisb@gmail.com}}

\author[0009-0005-8764-2608]{Sean McGuire}
\affil{LSST Interdisciplinary Network for Collaboration and Computing Frameworks, 933 N. Cherry Avenue, Tucson AZ 85721, USA}
\affil{The McWilliams Center for Cosmology \& Astrophysics, Department of Physics, Carnegie Mellon University, Pittsburgh, PA 15213, USA}
\foraas{\email{seanmcgu@andrew.cmu.edu}}

\author{Ian~E. Ordenes}
\affil{NSF NOIRLab/NSF–DOE Vera C. Rubin Observatory HQ, 950 N. Cherry Ave., Tucson, AZ 85719, USA}
\foraas{\email{iordenes@lsst.org}}

\author[0000-0002-4753-3387]{Maria~T. Patterson}
\affil{DiRAC Institute and the Department of Astronomy, University of Washington, 3910 15th Ave NE, Seattle, WA 98195, USA}
\foraas{\email{maria.t.patterson@gmail.com}}

\author[0000-0002-2598-0514]{Andrés~A. Plazas~Malagón}
\affil{SLAC National Accelerator Laboratory, Menlo Park, CA 94025, USA}
\affil{Kavli Institute for Particle Astrophysics \& Cosmology, P. O. Box
2450, Stanford University,Stanford, CA 94305,USA}
\foraas{\email{plazas@stanford.edu}}

\author{Yongqiang Qiu}
\affil{SLAC National Accelerator Laboratory, Menlo Park, CA 94025, USA}
\foraas{\email{qiu@slac.stanford.edu}}

\author[0000-0002-1557-3560]{Bruno Quint}
\affil{NSF NOIRLab/NSF–DOE Vera C. Rubin Observatory HQ, 950 N. Cherry Ave., Tucson, AZ 85719, USA}
\foraas{\email{bquint@lsst.org}}

\author[0000-0002-6975-827X]{Veljko Radeka}
\affil{Brookhaven National Laboratory, Upton, NY 11973, USA}
\foraas{\email{radeka@bnl.gov}}

\author{Wouter van Reeven}
\affil{Association of Universities for Research in Astronomy, 1331 Pennsylvania Ave. NW, Washington, DC 20004, USA}
\foraas{\email{wvanreeven@lsst.org}}

\author[0000-0001-8239-3079]{Vincent~J. Riot}
\affil{Lawrence Livermore National Laboratory, 7000 East Avenue, Livermore, CA 94550, USA}
\foraas{\email{riot1@llnl.gov}}

\author[0000-0002-8687-0669]{Bruno~O. Sánchez}
\affil{Aix-Marseille Université, CNRS/IN2P3, CPPM, Marseille, France}
\foraas{\email{bsanchez@cppm.in2p3.fr}}

\author[0000-0002-9238-9521]{David Sanmartim}
\affil{Association of Universities for Research in Astronomy, 1331 Pennsylvania Ave. NW, Washington, DC 20004, USA}
\affiliation{NSF NOIRLab/NSF–DOE Vera C. Rubin Observatory HQ, 950 N. Cherry Ave., Tucson, AZ 85719, USA}
\foraas{\email{dsanmartim@lsst.org}}

\author[0000-0001-9348-0290]{Jacques Sebag}
\affil{NSF NOIRLab/NSF–DOE Vera C. Rubin Observatory HQ, 950 N. Cherry Ave., Tucson, AZ 85719, USA}
\foraas{\email{jsebag@lsst.org}}

\author[0009-0000-6778-7168]{Alysha Shugart}
\affil{NSF NOIRLab/NSF–DOE Vera C. Rubin Observatory HQ, 950 N. Cherry Ave., Tucson, AZ 85719, USA}
\foraas{\email{alysha.shugart@noirlab.edu}}

\author[0000-0001-8708-251X]{Ian~S. Sullivan}
\affil{DiRAC Institute and the Department of Astronomy, University of Washington, 3910 15th Ave NE, Seattle, WA 98195, USA}
\foraas{\email{sullii@uw.edu}}

\author[0000-0001-6268-1882]{Dan~S. Taranu}
\affil{Department of Astrophysical Sciences, Princeton University, Princeton, NJ 08544, USA}
\foraas{\email{dtaranu@astro.princeton.edu}}

\author[0000-0002-3205-2484]{Elana~K. Urbach}
\affil{Department of Astronomy, Harvard University, 60 Garden St., Cambridge, MA 02138, USA}
\foraas{\email{eurbach@g.harvard.edu}}

\author[0000-0001-7113-1233]{W.~M.~Wood-Vasey}
\affiliation{
        Pittsburgh Particle Physics, Astrophysics, and Cosmology Center (PITT PACC).
        Physics and Astronomy Department, University of Pittsburgh,
        Pittsburgh, PA 15260, USA
}
\foraas{\email{wmwv@pitt.edu}}

\author[0000-0002-5596-198X]{Tianqing Zhang}
\affil{LSST Interdisciplinary Network for Collaboration and Computing Frameworks, 933 N. Cherry Avenue, Tucson AZ 85721, USA}
\affil{
        Pittsburgh Particle Physics, Astrophysics, and Cosmology Center (PITT PACC).
        Physics and Astronomy Department, University of Pittsburgh,
        Pittsburgh, PA 15260, USA
}
\foraas{\email{tq.zhang@pitt.edu}}

\begin{abstract}


The Vera C.~Rubin Observatory recently released Data Preview 1 (DP1) in advance of the upcoming Legacy Survey of Space and Time (LSST), which will enable boundless discoveries in time-domain astronomy over the next ten years.
DP1 provides an ideal sandbox for validating innovative data analysis approaches for the LSST mission, whose scale challenges established software infrastructure paradigms.
This note presents a pair of such pipelines for variability-finding using powerful software infrastructure suited to LSST data, namely the HATS (Hierarchical Adaptive Tiling Scheme) format and the LSDB framework, developed by the LSST Interdisciplinary Network for Collaboration and Computing (LINCC) Frameworks team.
This article presents: a pair of variability-finding pipelines built on LSDB; the HATS catalog of DP1 data; and preliminary results of detected variable objects, two of which are novel discoveries.


\end{abstract}

\foraas{\keywords{\uat{Transient detection}{1957} --- \uat{Eclipsing binary stars}{444} --- \uat{Short period variable stars}{1453} --- \uat{Light curves}{918} --- \uat{Catalogs}{205}}}

\section{Introduction}\label{sec:intro}


The Vera C.~Rubin Observatory Legacy Survey of Space and Time \citep[LSST:][]{2009arXiv0912.0201L,2019ApJ...873..111I} is a ten-year survey starting in 2025, with a goal of expanding our knowledge of the Solar System, the Milky Way, extragalactic astrophysics and cosmology, and understanding the time-varying sky.  The observing strategy has been defined through an iterative process with extensive community input \citep{2022ApJS..258....1B} to ensure that a wide variety of time-varying phenomena can be explored by LSST, while also leaving open opportunities for completely new discoveries.

Data Preview~1 \citep[DP1;][]{guy_2025_15558559} is a preliminary data release showing the capabilities of the observatory and allowing scientists to prepare their pipelines for future data releases, using data gathered with the Commissioning Camera (LSSTComCam; \citealt{SLAC_National_Accelerator_Laboratory2024-wv}).
DP1 was released to LSST data right holders on June 30, 2025 (\citealt{RTN-095}, \dataset[DP1 dataset]{\doi{10.71929/rubin/2570308}}).
The data release consists of calibrated, coadded and difference image analysis (DIA) images and catalog products, including both ``static" and time-domain photometry.

DP1 opens a door to the time-domain studies that LSST will enable in the near future.
LSSTComCam used the same six $ugrizy$ filters and provided the same single-exposure depth as the larger LSST Camera (LSSTCam) will provide.
This allows the community to validate that the data is suitable for time-domain analyses and to prepare their software for those analyses.  Given the scale and complexity of LSST, the Data Previews offer invaluable opportunities to prepare for LSST.

This paper presents a preliminary analysis performed by the LSST Interdisciplinary Network for Collaboration and Computing (LINCC) Frameworks\footnote{\url{https://lsstdiscoveryalliance.org/programs/lincc-frameworks/}} team.
The source code is available online, to enable others to reproduce and extend our analyses.\footnote{\url{https://github.com/lsst-sitcom/linccf}}  We describe the first time domain analysis of data from Rubin Observatory using software from LINCC Frameworks, in hopes of exploring the opportunities for discovery with data from Rubin.

We begin with a description of the input DP1 dataset and HATS version of the catalogs in Sec.~\ref{sec:data}.  We then outline our analysis and results in Sec.~\ref{sec:results}, including a comparison with other datasets where possible, and conclude in Sec.~\ref{sec:conclusion}.

\section{Data}\label{sec:data}

LSST~DP1 contains \fields{}~observation fields covering $\sim 15$~sq.~deg. in total.
The DP1 is composed from 1791 individual exposures, in one of $ugrizy$ filters, with field of view of $40' \times 40'$.
The cadence of this data is different from that of the future main survey: 
each field was observed multiple times per night, typically with up to three different filters, as opposed to observations every few days with a pair of filters per night.
This enhanced cadence allows us to run multi-band periodicity analysis for timescales as short as an hour.

Our afterburner pipeline transforms the DP1 catalog data products into science-ready catalogs in the HATS (Hierarchical Adaptive Tiling Scheme) format \citep{2025arXiv250102103C}.
HATS is designed to store large astronomy catalogs with parquet files and offers high-level metadata files for exploring the scope of the dataset.
Rubin data-rights holders can access these catalogs via the Rubin Science Platform or Independent Data Access Centers\footnote{Please find the catalog access instructions at \url{https://lsdb.io/dp1}.}.

The survey-provided catalogs are modified for ease of scientific exploration in the following ways:
\begin{itemize}
 \item time domain photometry is nested within the same parquet files as the ``static" information about the objects (or DIA objects);
 \item we remove any rows without valid celestial coordinates;
 \item we perform an offline join to append columns to that correspond to Rubin's Butler dimensions (e.g. \texttt{tract}, \texttt{patch}, \texttt{band}) \citep{2022SPIE12189E..11J};
 \item for all point spread function (PSF) \texttt{flux} (and \texttt{fluxErr}) columns, we calculate the corresponding AB magnitude and add \texttt{mag} (and \texttt{magErr}) values;
 \item we perform an offline join to the original visits to retrieve and append the midpoint of exposure time \texttt{midpointMjdTai} for all source types.
\end{itemize}

We provide both an \texttt{object} and \texttt{diaObject} table wherein each type of object has associated forced photometry light curves, based on both direct science images and difference images.
The \texttt{object} table has \TotalObjects{} rows and the total of \TotalObjectForcedSources{} forced sources, while \texttt{diaObject} table has \TotalDiaObjects{} rows, \TotalDiaSources{} DIA sources, and \TotalDiaObjectForcedSources{} forced sources.

\section{Methods and Results}\label{sec:results}

We use LSDB\footnote{\url{https://lsdb.io}} to leverage the HATS-formatted DP1 catalogs. 
LSDB is a Python framework for large-scale astronomical catalog analysis, which enables parallel and out-of-memory computation using both built-in and user-provided functions~\citep{2025arXiv250102103C}.
LSDB is based on \texttt{Pandas} and \texttt{Dask} and represents an astronomical catalog as a collection of lazily-loaded data frames, each containing objects from a distinct region of the sky.
Since LSDB uses HATS as the data storage format, they both share the same sky tiling schema, HEALPix \citep{2005ApJ...622..759G}, distributing objects homogeneously over individual data frames.
LSDB enables time-domain analysis using \texttt{nested-pandas}{\footnote{\url{https://github.com/lincc-frameworks/nested-pandas}}}.
\texttt{nested-pandas} packs each light curve as a single data-frame element, which simplifies distributed execution and user's access to the photometric data.

Using LSDB, we developed two search pipelines: 
flare detection, and periodicity analysis.
Both pipelines use \texttt{diaObject} table objects with at least five observations with signal-to-noise ratio larger than three and at least eight observations in total.

\subsection{Transients}

The flare detection pipeline performs per-photometric-filter parametric fits for both a constant function and a Bazin function with exponential asymptotics both for the rising and the falling stages of a flare \citep{2009A&A...499..653B}:
\begin{equation}\label{eq:bazin}
   f(t) = A \frac{ \mathrm{e}^{ -(t-t_0)/\tau_\mathrm{fall} } }{ 1 + \mathrm{e}^{ -(t - t_0) / \tau_\mathrm{rise} } } + B,
\end{equation}
where $f(t)$ is the bandflux as a function of time, $A$ and $B$ are scale and offset in the bandflux units, $t_0$ is the reference time, and $\tau_\mathrm{fall}$ and $\tau_\mathrm{rise}$ are the characteristic rising and falling times.

We then select single-filter light curves with reduced $\chi^{2}_{c}$ of the constant fit larger than one, reduced  $\chi^{2}_{B}$ of the Bazin fit smaller than ten, and the reduced $\chi^{2}_{c} / \chi^{2}_{B}$ ratio smaller than three.
We then perform a visual inspection to identify flaring objects.

Here we present Rubin-only detections of flaring objects, a mix of previously classified and unclassified objects: 
two quasars\footnote{While quasars are not transients, for the short duration of these observations their approximately linear light curves are well-fit by a Bazin function, so they were identified with the approach in this section.}, two cataclysmic variables, a transient, and an M-dwarf flare. 
Visual inspection also revealed false detections, typically due to photometric pipeline problems that may improve in future versions of the processing.

{
The light curves of the detected quasars are shown in Figure~\ref{fig:quasars}.
The left panel shows radio-loud quasar J033227.0$-$274105 / GS-44b at $\alpha=53.11256$, $\delta=-27.68482$ \citep{2006A&A...455..773V, 2019ApJ...875...80G}, with DP1 $m_g \simeq 19.6$, 
whose light curve demonstrates a linear decay. 
The right panel shows a quasar at $\alpha= 52.69666$, $\delta = -27.98802$ \citep{2010ApJ...711..928C}, with DP1 $m_g \simeq 20.4$,
whose light curve rises before MJD $\sim 60630$.
Both objects, not being transients, still have a good Bazin function fit for $r$ band because of the near-linear decay.
\begin{figure*}
    \centering
    \includegraphics[width=0.45\linewidth]{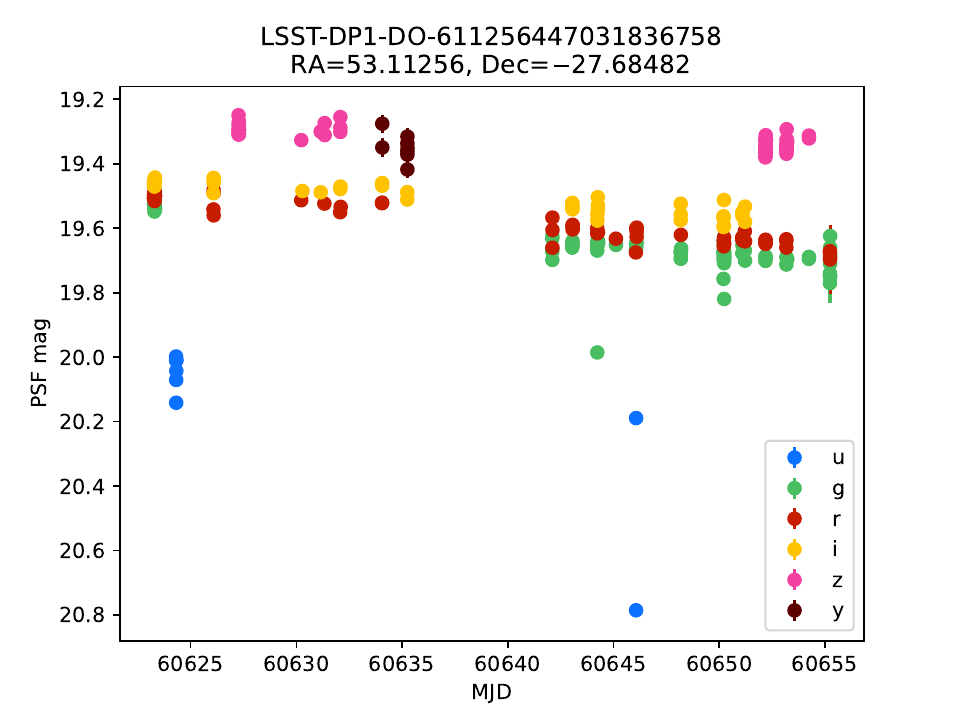}
    \includegraphics[width=0.45\linewidth]{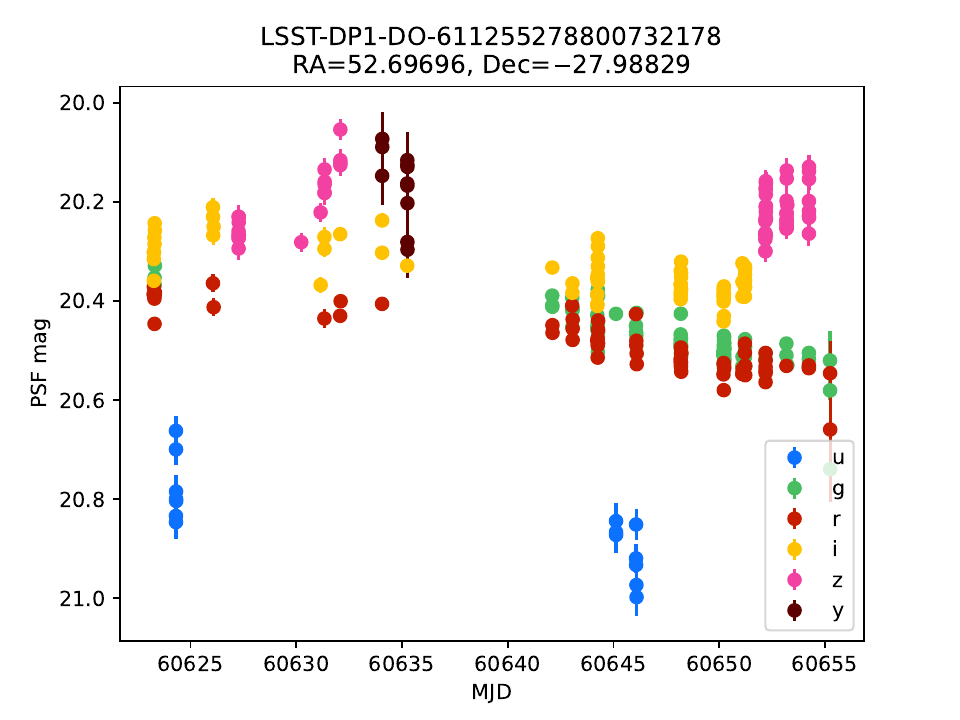}
    \caption{
    {Quasars found in DP1 with our transient search pipeline. 
    Left: LSST-DP1-DO-611256447031836758.
    Right: LSST-DP1-DO-611255278800732178. 
    Observations with magnitude error larger than 0.2 are not shown. As shown, in most bands the light curves are approximately linear, which is why they can be well fit with Eq.~\ref{eq:bazin}.}
    }
    \label{fig:quasars}
\end{figure*}
}

{
We find outbursts of two previously classified cataclysmic variables \citep[for a review, see][]{1995cvs..book.....W}, shown in Figure~\ref{fig:dwarf-novae}.
CRTS~J033349.8$-$282244 ($\alpha=53.45743$, $\delta=-28.37869856$) \citep{2014MNRAS.441.1186D} has $m_g \simeq 18.96$ while quiescent and $m_g(\mathrm{MJD}\ 60644.257) = 15.15$ at peak.
ZTF19acecitx ($\alpha=95.45970$, $\delta=-24.72973$) was assigned a dubious AM~CVn classification when first discovered \citep{2021AJ....162..113V}, and in DP1 it demonstrates a flare spanning the whole observational range, peaking at $m_g(\mathrm{MJD}\ 60646.333) = 19.43$.
\begin{figure*}
    \centering
    \includegraphics[width=0.45\linewidth]{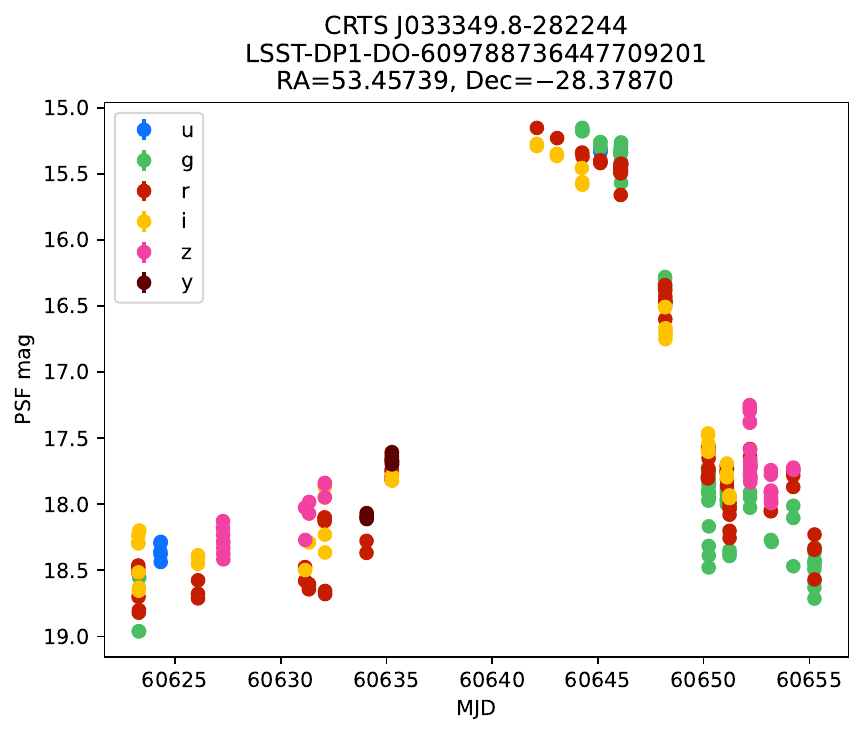}
    \includegraphics[width=0.45\linewidth]{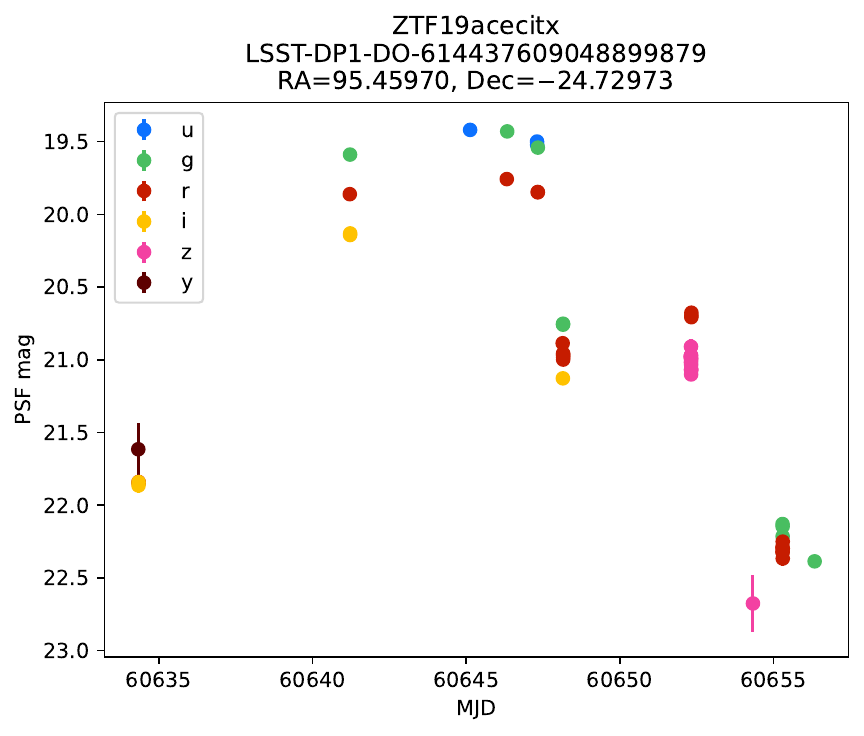}
    \caption{
    {Cataclysmic variable outbursts detected in DP1. 
    Left: LSST-DP1-DO-609788736447709201 / CRTS~J033349.8$-$282244. 
    Right: LSST-DP1-DO-614437609048899879 / ZTF19acecitx.  In both cases, the rising and falling parts of the light curve are observed in multiple filters.}
    }
    \label{fig:dwarf-novae}
\end{figure*}
}

We have {also} found the transient AT2024ahyy ($\alpha = 52.89263$, $\delta=-28.41263$), discovered by the Young Supernova Experiment \citep[YSE;][]{2019ATel13330....1J} on Dark Energy Camera \citep[DECam;][]{2008SPIE.7014E..0ED} with a single $i$-band detection $m_i(\mathrm{MJD}\ 60646.816) = 23.17\pm0.24$ \citep{2022TNSAN..24....1R,2025TNSTR.975....1M}.
After we adjust the DP1 DIA light curve so the pre-transient fluxes are zero, we find that it has 111 observations with signal-to-noise ratio larger than three.
The first detection has ``direct" (non-DIA) $r$-band magnitude $m_r(\mathrm{MJD}\ 60646.062) = 22.56 \pm 0.05$.
{
The light curve of the object is shown in Figure~\ref{fig:transient}; as shown, the single DECam observation is consistent with the many observations available in DP1.
\begin{figure*}
    \centering
    \includegraphics[height=0.35\linewidth]{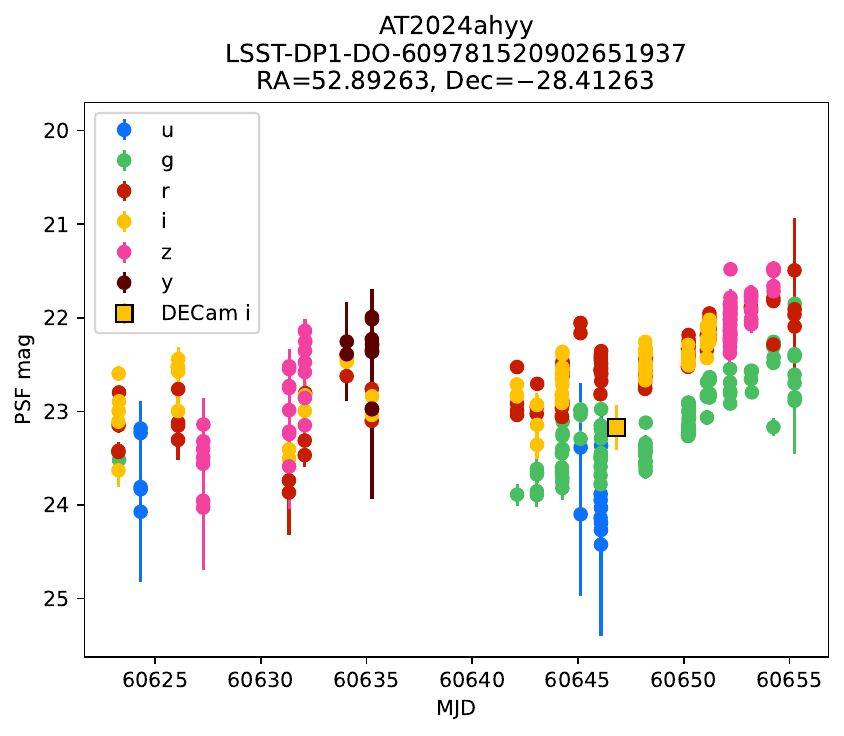}
    \includegraphics[height=0.35\linewidth]{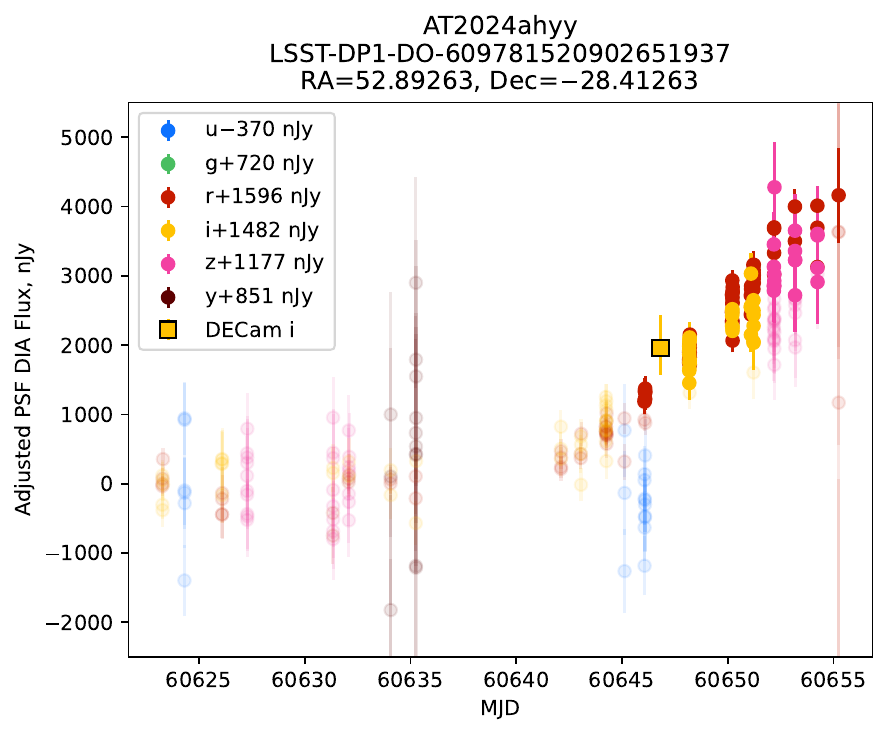}
    \caption{
    {
    DP1 and YSE DECam photometry of LSST-DP1-DO-609781520902651937 / AT2024ahyy. 
    Left: AB magnitude.
    Right: bandfluxes, with DP1 difference bandflux pre-transient photometry shifted to zero before MJD 60640, and DECam AB magnitudes being converted to nJy.
    Semi-transparent points mark measurements with signal-to-noise ratio smaller than three, where the noise term includes the uncertainty in the pre-transient flux estimation.  The rise in the light curve can be clearly detected in multiple filters.
    }
    }
    \label{fig:transient}
\end{figure*}
}

\begin{figure*}
    \centering
    \includegraphics[width=0.45\linewidth]{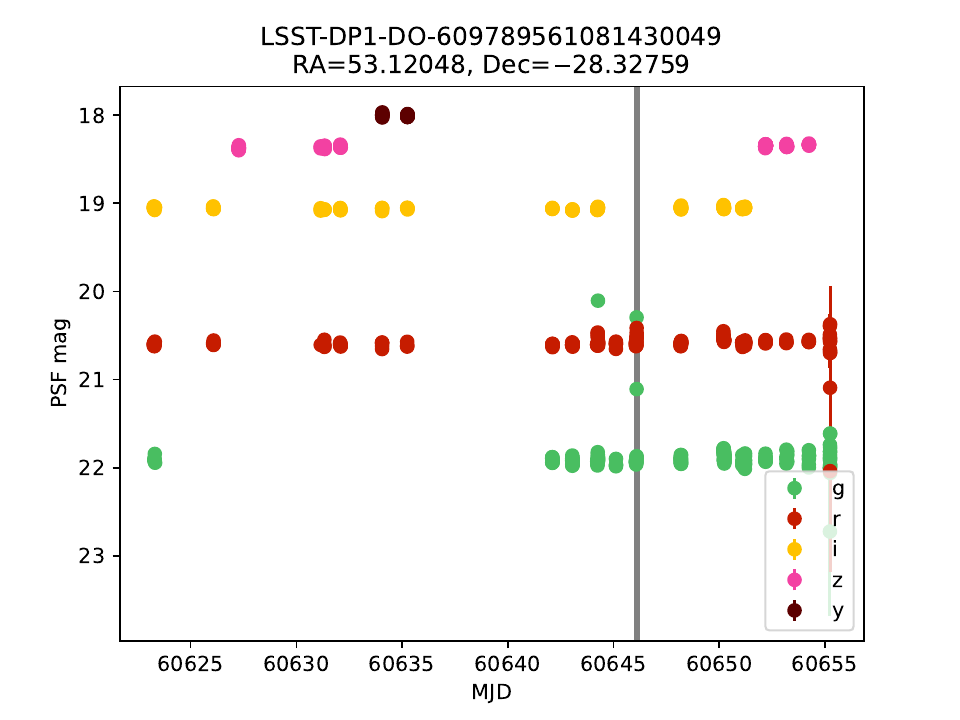}
    \includegraphics[width=0.45\linewidth]{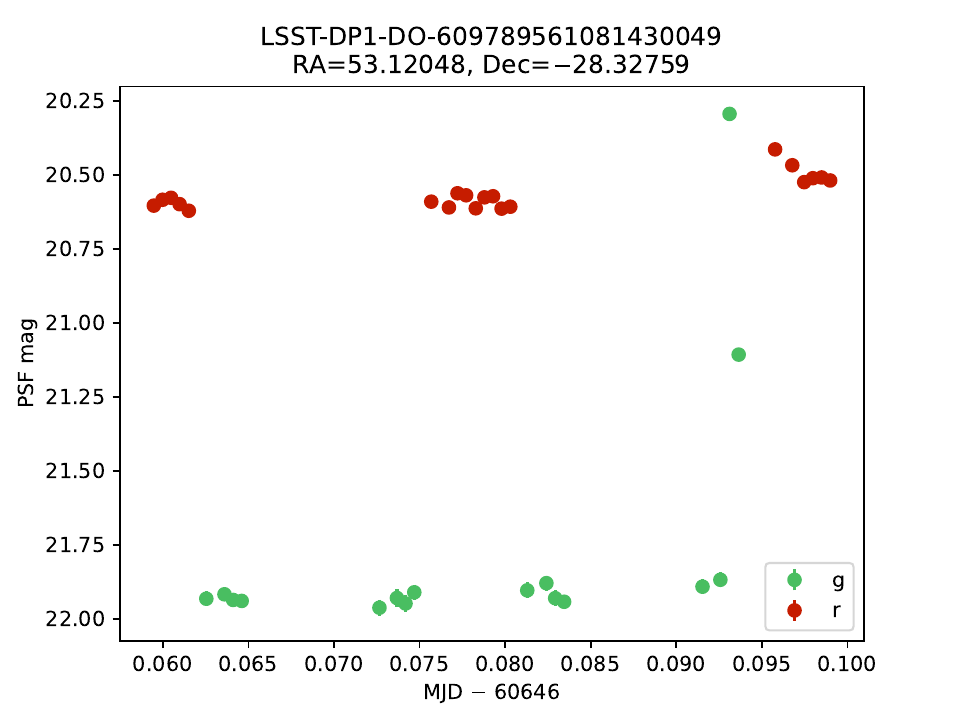}
    \caption{
    M dwarf flare LSST-DP1-DO-609789561081430049. Left panel: full light curve, the detected flare is shown with a gray line at MJD 60646. Right panel: an hour of observations at MJD 60646 covering the flare.}
    \label{fig:m-dwarf-flare}
\end{figure*}

Finally, our transient-detection pipeline discovered a short transient, most likely an M-dwarf flare, shown in Figure~\ref{fig:m-dwarf-flare}.
The colors of the object in quiescence correspond to those of an M4 dwarf ($\alpha=53.12048$, $\delta=-28.32759$): 
$r-i \simeq 1.53$ and $i-z \simeq 0.71$ \citep{2009AJ....138..633K}.
Quiescent at $m_g = 21.90 \pm 0.05$, the flare peaked at $m_g(\mathrm{MJD}\ 60646.0931) = 20.29 \pm 0.01$ and appears in $g$ and $r$ bands.
It is hard to tell the total duration of the flare because the tail is not fully captured, but the eight observations during the flare span about 10 minutes.
The light curve has one additional outlier point at $m_g(\mathrm{MJD}\ 60644.2664) = 20.10 \pm 0.01$, which probably corresponds to another flare.

\subsection{Periodic variability}\label{sec:periodic}

As inputs to the periodicity pipeline, we selected light curves with at least 50 total observations and at least 30 observations in any of the $griz$ bands.
Because of the cadence properties of the DP1 dataset, we focus our search on periods from 5 minutes to 12 hours, which are potentially affected by light travel time differences due to the Earth's orbital movement, so we convert the observational mid-exposure
MJD
provided by the DP1 to heliocentric MJD.
We calculate the Lomb---Scargle periodogram \citep{1976Ap&SS..39..447L,1982ApJ...263..835S} using \texttt{astropy} for each of the $griz$ filters, and find the peaks of the periodogram.
We also exclude spurious periods around 1/4 and 1/3 day, which are harmonics of the strong 1-day periodicity arising from regular daily gaps in ground-based observations.
Then we select objects where the best periods in at least two bands are within 0.1\% and filter out peaks for which the minimum false alarm probability exceeds $10^{-10}$ \citep{2008MNRAS.385.1279B}.
The pipeline reveals multiple periodic variables, mostly RR Lyrae variables and eclipsing variables with ellipsoidal components.

{Figure~\ref{fig:periodics} shows a few examples of DP1-detected periodic variables previously classified by the Catalina Sky Survey \citep{2017MNRAS.469.3688D} and Gaia \citep{2023A&A...674A..18C}.
For these plots we estimated the periods again, this time using multi-band periodogram \citep{2015ApJ...812...18V} applied to photometry combined from all available bands of Rubin and other surveys: Catalina, Gaia, and ATLAS \citep{2018PASP..130f4505T,2018AJ....156..241H,2021TNSAN...7....1S}.
We believe that these estimates may improve previously known periods because of the larger time baseline.
\begin{figure*}
    \centering
    \includegraphics[width=0.45\linewidth]{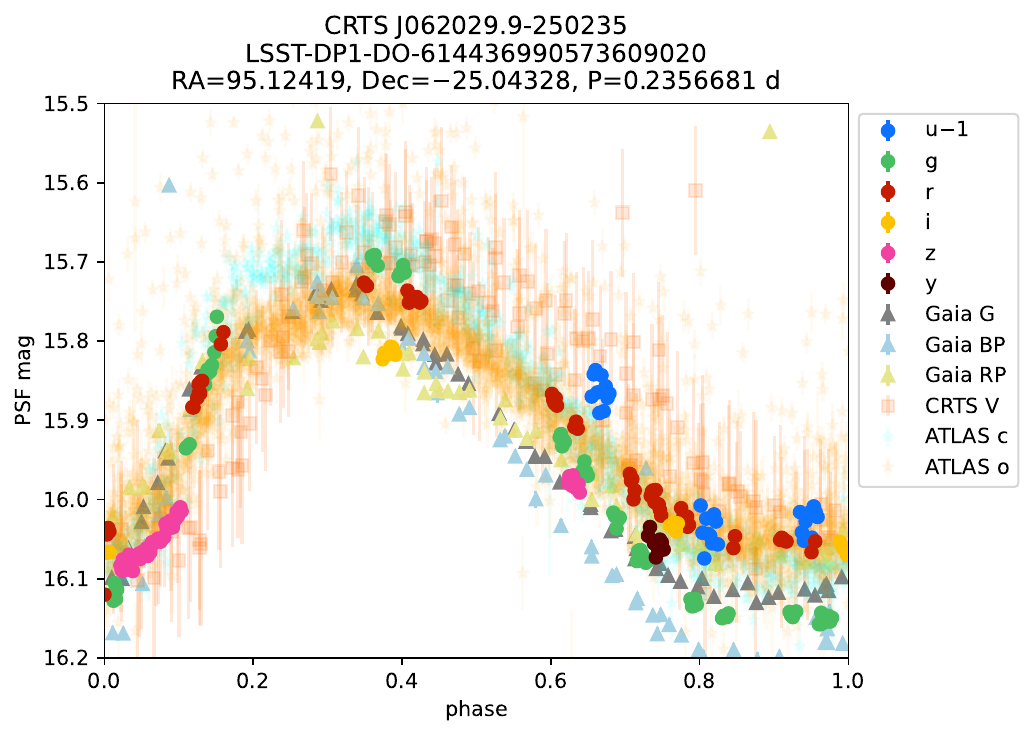}
    \includegraphics[width=0.45\linewidth]{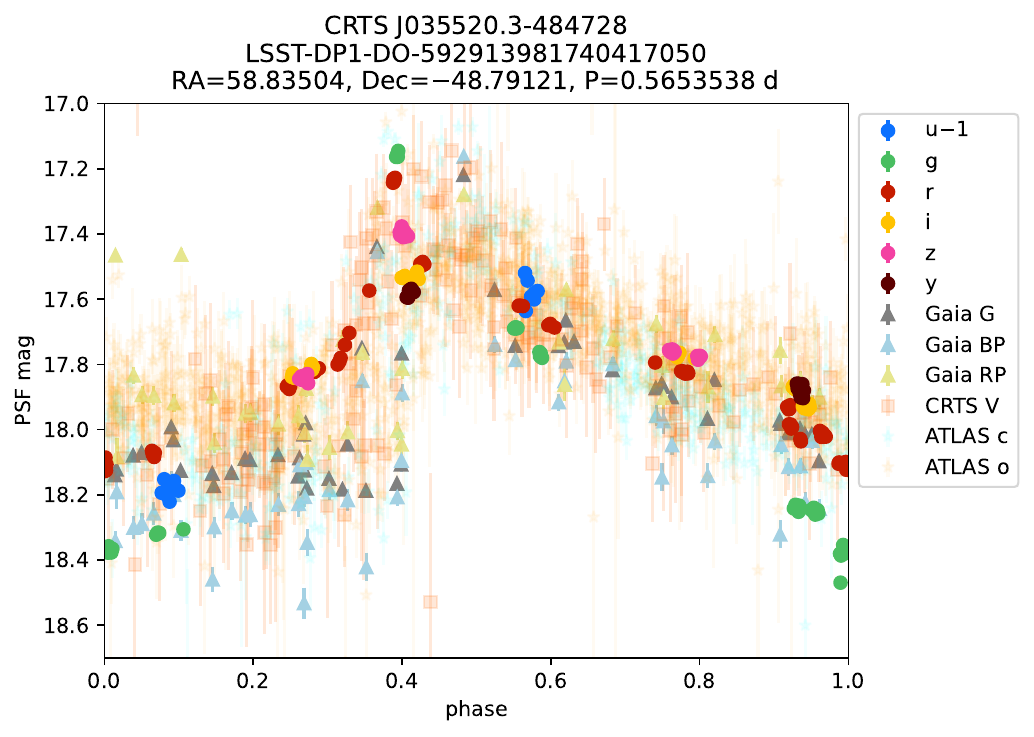}
    \includegraphics[width=0.45\linewidth]{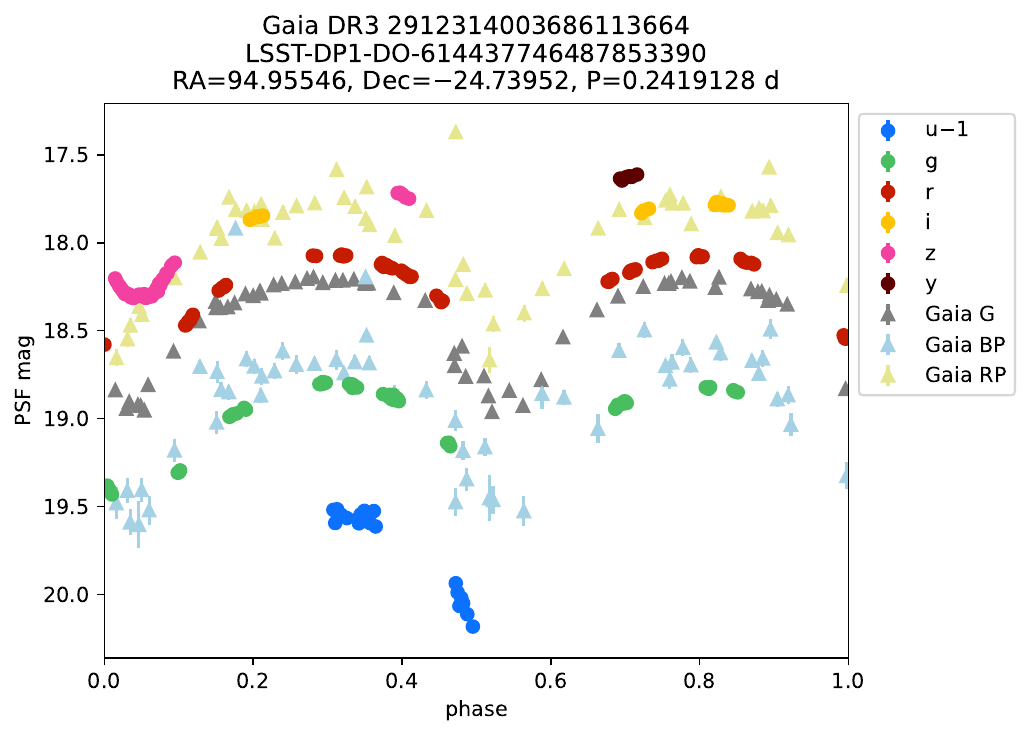}
    \includegraphics[width=0.45\linewidth]{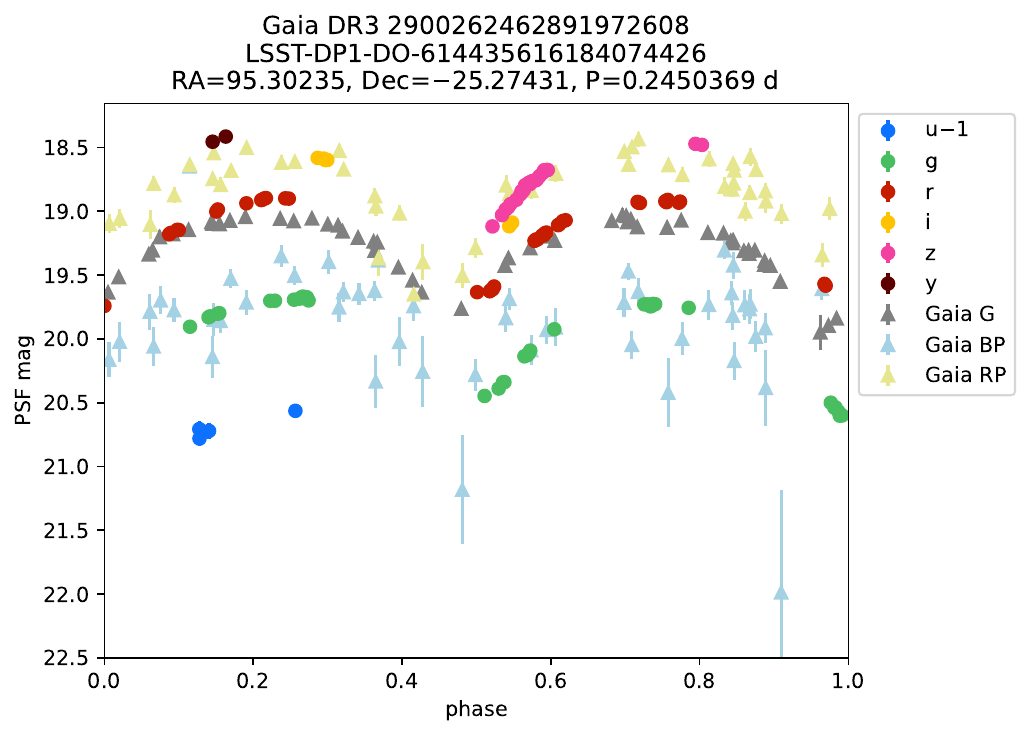}
    \includegraphics[width=0.45\linewidth]{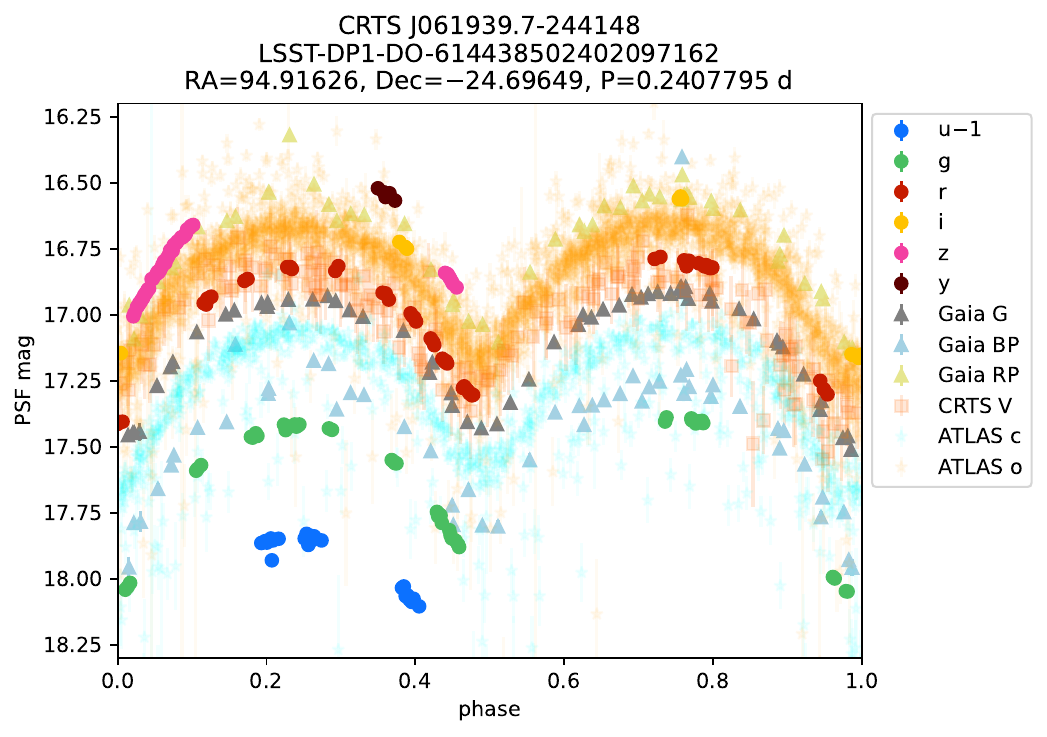}
    \caption{
    {Examples of periodic variables in the DP1 data enriched with Catalina Sky Survey, Gaia and ATLAS light curves. 
    Top row: RR~Lyrae.
    Central and bottom rows: eclipsing binaries.
    Note that DP1's $u$-band photometry is shifted by one magnitude.}
    }
    \label{fig:periodics}
\end{figure*}
}

\begin{figure}
    \centering
    \includegraphics[width=0.45\linewidth]{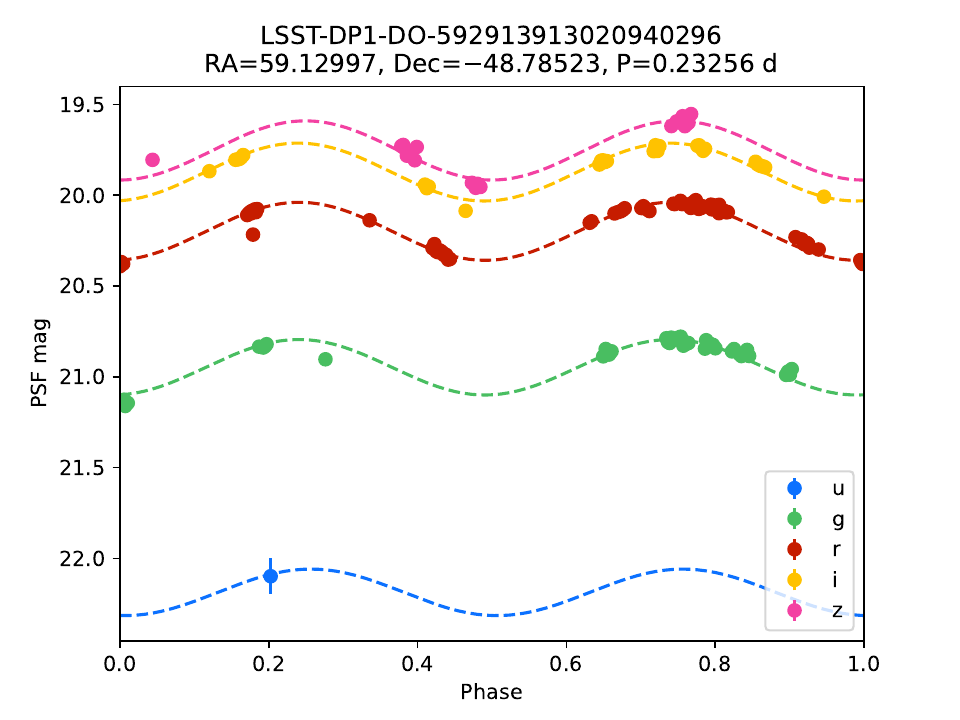}
    \caption{Period-folded eclipsing binary LSST-DP1-DO-592913913020940296, with a half-period sine fit (dashed line), period $P=0.23256$ days.}
    \label{fig:eclipsing}
\end{figure}

One of the periodic variables found was previously unclassified and is presented in Figure~\ref{fig:eclipsing}, $\alpha = 59.12997$, $\delta = -48.78523$.
We fit LSST's $griz$-, Gaia~DR3 $G$-, $RP$-, $BP$-, and DES~DR2 $Y$-band photometry \citep{2023A&A...674A...1G,2021ApJS..255...20A} to the black-body model.
This gives us an estimate of the color temperature and the solid angle, the least can be converted to distance if the stellar radius is known. The effective temperature is $4700 \pm 70$~K, and the distance is $(R/R_\odot)^2 \times 10.0 \pm 0.7$~kpc $\simeq 5.8 \pm 0.6$~kpc, where $R/R_\odot \simeq 0.73$ is the radius in solar units, obtained by linear interpolation of the dwarf star temperature–radius relation \citep{2013ApJS..208....9P}.
{This distance estimate is close to the photo-geometrical distance of $7.8^{+1.4}_{-1.0}$~kpc given by \cite{2021AJ....161..147B}.}
Our characterization of the spectral and luminosity classification, period, and amplitude values suggest that the object is most likely to be a W~Ursae Majoris eclipsing binary, with the primary component to be a K-dwarf.

{
\section{Conclusion}\label{sec:conclusion}

In this paper, we presented the results of a time series analysis of Rubin Data Preview 1 using the LSDB framework.
We ran search pipelines to identify light curves exhibiting two types of behavior: exponential rise and decay, and periodic variability.
While we found multiple objects that had previously been detected and classified using photometric data from earlier surveys,
the depth of the DP1 LSSTComCam images and the efficiency of LSDB enabled us to identify an M-dwarf flare and previously unclassified eclipsing binary and variable objects as well.  

Our analysis serves as an initial proof of concept for use of LSDB to identify scientifically-interesting time-varying phenomena in data from Rubin Observatory. The results were quite promising given the primary limitations in this analysis -- for example, the limited area and total duration of the observations with LSSTComCam that were released in DP1. We anticipate a considerably richer time domain dataset with the upcoming Data Preview 2 (DP2) based on all science-grade commissioning data with LSSTCam \citep{guy_2025_15558559}.  The transition from DP1 to DP2 will also enable a scaling test for LSDB as an intermediate step before the first Data Release for LSST.

Finally, we release the HATS versions of the DP1 catalogs used for this analysis.
The HATS catalogs are available for LSST data rights holders via the Rubin Science Platform\footnote{\url{https://data.lsst.cloud}} and Rubin Independent Data Access Centers\footnote{\url{https://rtn-060.lsst.io}}, enabling at-scale analysis of the data with LSDB.
}



\begin{acknowledgments}
LINCC Frameworks is supported by Schmidt Sciences.
This publication is based in part on proprietary Rubin Observatory data and was prepared in accordance with the Rubin Observatory data rights and access policies.
This material is based upon work supported in part by the National Science Foundation through Cooperative Agreements AST-1258333 and AST-2241526 and Cooperative Support Agreements AST-1202910 and 2211468 managed by the Association of Universities for Research in Astronomy (AURA), and the Department of Energy under Contract No. DE-AC02-76SF00515 with the SLAC National Accelerator Laboratory managed by Stanford University. Additional Rubin Observatory funding comes from private donations, grants to universities, and in-kind support from LSST-DA Institutional Members. This paper is based upon work supported by the National Science Foundation under Grant No. AST-2003196.
\end{acknowledgments}

\facilities{Rubin:Simonyi (LSSTComCam), ATLAS, Blanco, Gaia.}

\software{
\texttt{astropy} \citep{2013A&A...558A..33A,2018AJ....156..123A,2022ApJ...935..167A},
{\texttt{Dask} \citep{matthew_rocklin-proc-scipy-2015},}
\texttt{LSDB} \citep{2025arXiv250102103C},
\texttt{light-curve} \citep{2021MNRAS.502.5147M}%
{, \texttt{Matplotlib} \citep{Hunter2007}, }%
{\texttt{Numpy} \citep{harris2020array}, }%
{\texttt{Pandas}
\citep{reback2020pandas}},
\texttt{Scipy} \citep{2020SciPy-NMeth}%
.
}

\bibliography{dp1-lsdb}{}
\bibliographystyle{aasjournalv7}

\end{document}